\documentclass[a4paper,11pt]{article}
\usepackage{pos}

\usepackage{enumitem}

\title{The Czech Particle Physics Project}

\author*[ab]{Andr\'e Sopczak}
\author[a]{Peter Žáčik}

\affiliation[a]{Institute of Experimental and Applied Physics,
  Czech  Technical University in Prague, Prague, Czechia}
\affiliation[b]{Ulrich Bonse Visiting Chair for Instrumentation,
TU Dortmund University, Dortmund, Germany}

\emailAdd{andre.sopczak@cern.ch}

\abstract{We present the status of the Czech Particle Physics Project (CPPP). The CPPP is a learning tool in masterclasses aimed at high school students (aged 15 to 18). The project is structured in modules. The first module is dedicated to the detection of an Axion-Like-Particle (ALP) using the ATLAS Forward Proton (AFP) detector. The second module focuses on the reconstruction of the Higgs boson mass using the Higgs boson golden channel with four leptons in the final state. 
The third and fourth modules are educational aids and 
sources for expert information. These web portals are dedicated to Higgs boson research and searches for Supersymmetry.
Two databases are created with more than 1000 relevant articles each, using the CERN Document Server API and web scraping methods.
The databases are automatically updated when new results on Higgs bosons or searches for Supersymmetric particles become available.
Using artificial intelligence and natural language processing, 
the articles are categorized according to properties of the results.
The modules are accessible at \url{https://cern.ch/cppp}.}

\FullConference{42nd International Conference on High Energy Physics (ICHEP2024)\\
18-24 July 2024\\
Prague, Czech Republic\\}

\begin{document}
\maketitle

\section{Introduction}
\vspace*{-4mm}
 Outreach is an important part of CERN's mission. An audience that is particularly important consists of high school students who might be interested in pursuing further studies in the
 Science, Technology, Engineering and Mathematics (STEM) fields. 
 A set of online materials is presented that can be used in 
 masterclasses to teach high school students (15 to 18 years old) about particle physics at CERN. 
 These tools are part of the wider Czech Particle Physics Project (CPPP). 
  
    Two modules have been developed, each having two main components: 
    an introduction which explains the basic physics involved and 
    a simulation which allows users
    to learn about physics research interactively.
    The first module takes students through the process of finding an Axion-Like Particle (ALP) using the ATLAS Forward Proton (AFP) detector. The second module shows students how to determine the mass of 
    the Higgs boson by studying the so-called golden channel 
    ($H \to ZZ \to 4e$).
Details of the implementation of the modules are given in Refs.~\cite{Zacik:2774895,ippog2021}.

Over the past decades, the search for the Higgs boson of the Standard Model and Higgs bosons in extended models have been at the forefront of research in particle physics. About twelve years ago the discovery of a Higgs boson with Standard Model properties was established.
The Higgs boson research has led to more than 1000 experimental publications on the Higgs boson search and the measurement of its properties. Many aspects of the Higgs boson research have been addressed, including the search for Higgs bosons beyond the Standard Model.
For the benefit of experts and the general public interested in this exciting field of research, a Higgs Boson Portal (HBP) is implemented as
part of the CPPP.
Recently, a new module for Supersymmetry searches has been added. The Supersymmetry Portal (SP) is structured similar to the HBP and also includes more than 1000 publications. 

The HBP and SP use 
Artificial Intelligence (AI) and
Natural Language Processing (NLP) for categorization of the Higgs boson and Supersymmetry publications in a user-accessible list with the following categorizations for the HBP:
Publication stage (preprint, journal accepted, published);
Year of public release;
Experiment;
Luminosity;
Higgs boson decay mode;
Higgs boson production mode;
Higgs boson model (Standard Model or Beyond Standard Model).
The detailed categorization 
of the production and decay modes 
follows the review~\cite{Sopczak:2020vrs}.
For the SP, specific  categories are: Final states with jets, leptons, photons; Displaced vertices / Long-lived particles;
R-parity violation; MSSM.
The SP structure follows the reviews~\cite{sopczak2014hunt,Sopczak:2303673}.

The database is automatically updated when new results on the Higgs boson or Supersymmetry searches become available, and
the components are deployed to CERN Web Services.
The HBP  also includes a visualisation of some developments of limits and precisions.
Following feasibility studies~\cite{Kupka:2722144,Demchenko:2753953},
implementation~\cite{Zacik:2774895} and 
presentation~\cite{ippog2021},
the HBP and SP are accessible on \url{https://higgs.web.cern.ch} and \url{https://susy.web.cern.ch}, respectively.

\begin{figure}[!hbtp]
    \vspace*{-8mm}
    \centering
    \includegraphics[width=\textwidth]{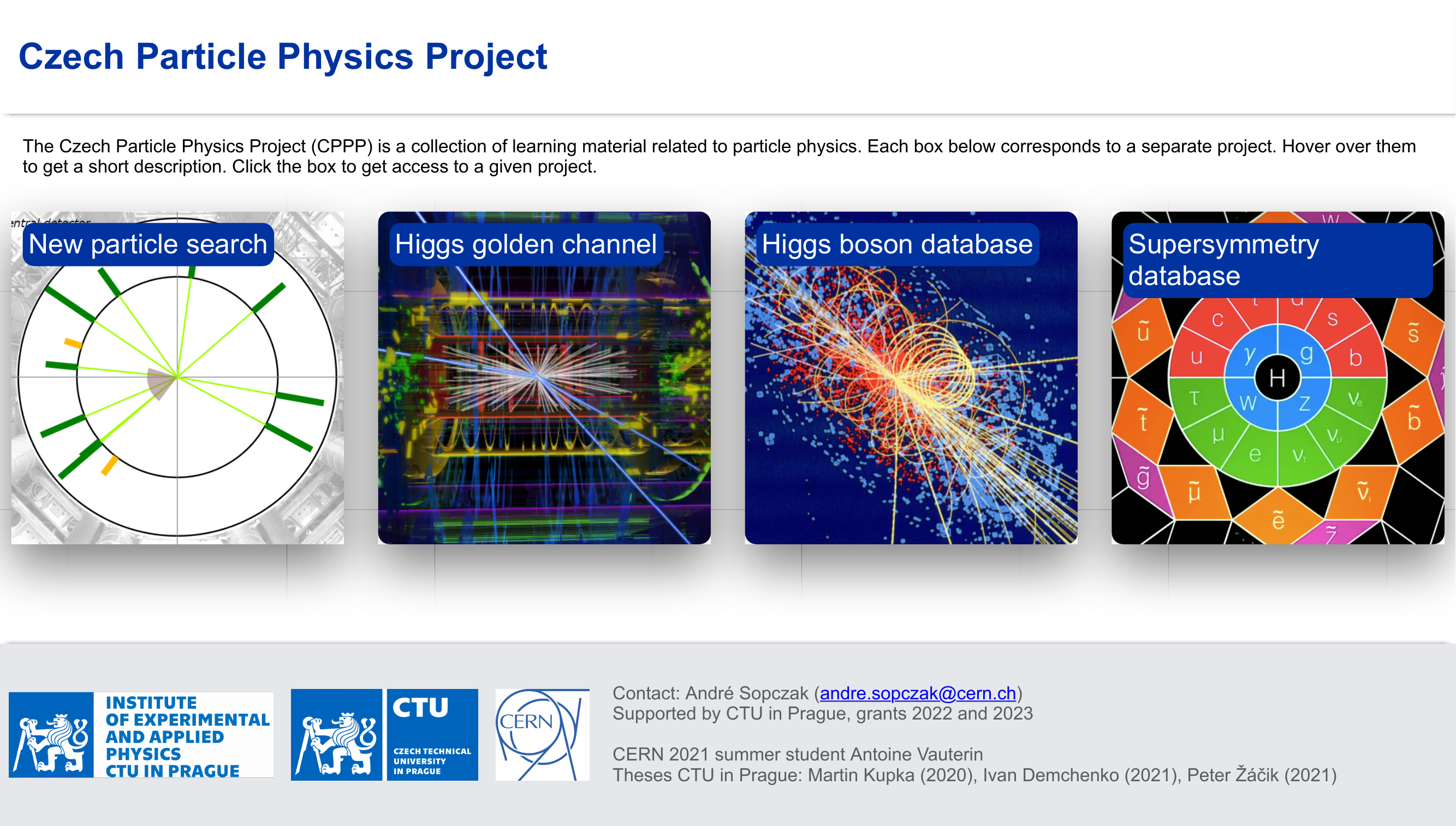}
    \vspace*{-7mm}
    \caption{Homepage of the Czech Particle Physics Project (CPPP).}
    \label{fig:homepage}
    \vspace*{-4mm}
\end{figure}

The CPPP 
is  a 
wide-ranging outreach project with a home page 
(Fig.~\ref{fig:homepage})  which lets the user easily access different modules.
Currently, it contains the ALP and Higgs boson golden channel modules, as well as modules that allow users 
to search databases of publications with automated update and categorization
related to the experimental Higgs boson 
research\footnote{Following a feasibility study by Martin Kupka~\cite{Kupka:2722144}, implemented by Peter Zacik~\cite{Zacik:2774895}.}
and searches for Supersymmetry\footnote{Following a 
feasibility study by 
Ivan Demchenko~\cite{Demchenko:2753953}, implemented by Peter Zacik.}.
The CPPP is expected to be expanded with new modules 
added to the home page in the structure
that is provided.

\section{Finding an ALP using the AFP detector}

\subsection{ALP and AFP}

Axion-Like-Particles (ALP) are potential dark matter candidates that are predicted by some extensions of the Standard Model~\cite{Baldenegro:2729326}. 
These new particles could be detected with the ATLAS detector at CERN.
The ATLAS Forward Proton (AFP) detector is used to enhance
the sensitivity for a discovery.
The AFP detector is located on either side of the ATLAS main detector 
and it is able to detect protons that come from the central $pp$ interaction point but are deviated from the axis of the beampipe. 
Two incoming protons could pass closely by each other 
instead of strongly interacting 
in which case they will interact 
through their electromagnetic fields and be deflected. 
This can produce a pair of photons. If ALPs exist the photons could scatter through an ALP ($\gamma \gamma \to ALP \to \gamma \gamma$), such that the scattered photons are detected in the central ATLAS detector while the deflected protons are detected in the AFP detector.
There are also  diphoton events without an ALP being involved 
which are referred to as  background events and 
can be largely removed by the two calculated energy loss values from 
the diphoton system and the deflected protons. 
If the scattering happened through an ALP the calculated 
energy loss should match. 
This criterion allows us to increase the ratio of 
signal-to-background events and thus isolate the signal better. 
The fact that the information from an additional detector 
enhances the sensitivity to find a new particle 
is an important learning aspect.

\subsection{Introduction page}

Before obtaining access to the simulation, 
the user reads through an introduction page which has five sections.
This introductory page contains all the information the user needs to appreciate the simulation. It should also stimulate further reading about CERN, LHC and the ATLAS detector.

\subsection{User experience}

The main components of the interactive event display are:

\begin{enumerate}[topsep=0pt,itemsep=-1ex,partopsep=1ex,parsep=1ex]
    \item Control panel: allows the user to interact with the simulation
    \item Event counters: counter for the total number of events fired and counter for auto-fired events
    \item Inner ATLAS central detector
    \item ATLAS central detector side view with AFP detectors on either side: animation of the protons being fired
    \item Energy loss matching histograms: the energy loss calculated from the AFP detectors in red and from the ATLAS central detector in green
    \item Invariant mass histograms: left histogram for all events, right histogram only for events where the energy loss values are matched (as decided by the user, or automatically in auto-fire mode)
    \item Access to admin page: password protected
\end{enumerate}

Once the user loads the simulation page, a succession of post-it notes on the screen will guide through the steps needed to run the simulation.

After going through the analysis process manually at least five times, the user unlocks the auto-fire feature. This allows the user to generate a large number of events automatically. In general, around 100 events are needed to observe a clear signal in the simulated data. 

Once the user has gone through the full auto-fire mode, a quiz pops-up which asks the user to locate the signal. If this is done correctly, an explanation is given and the user is free to continue with the simulation. If a wrong answer is given, the user is encouraged to continue firing more events and collect more statistics.

\subsection{Admin page}

The website has a password protected admin page. This page allows an administrator to modify the parameters of the simulation. 
Changes on this page are sent to the server and are applied globally, 
thus anyone on the website will have the new parameters.
A possible extension of the project could be that individual user groups can define their own Admin settings.

\section{Higgs boson and Supersymmetry modules}

The Higgs boson golden channel is one of the research modes of the Higgs boson~\cite{Aaboud:2277731}. 
In this channel, the Higgs boson decays into a pair of $Z$ bosons (real or virtual). Each $Z$ boson then decays into a pair of light leptons (electron or muon).
The signal for the Higgs boson is a signature with four leptons 
with an invariant mass equal to the Higgs boson mass. 
The main background to this signal is the production of di-$Z$
bosons directly from the proton collision decaying to leptons.
For simplicity, the module focuses only on the four-electron final state.

This process has been observed with the ATLAS detector~\cite{Aaboud:2277731}. 
The signal peaks at the Higgs boson mass (125 GeV). 
The background covers a wide range of invariant masses and 
peaks at the $Z$ boson mass around 91~GeV.
In this module, the user learns how to identify 
the four leptons by applying a selection 
on the lepton transverse momentum, 
and thus the user will reproduce the ATLAS distribution 
with the $Z$ and Higgs boson mass peaks.

\begin{figure}[h]
        \centering
    \includegraphics[width=\textwidth]{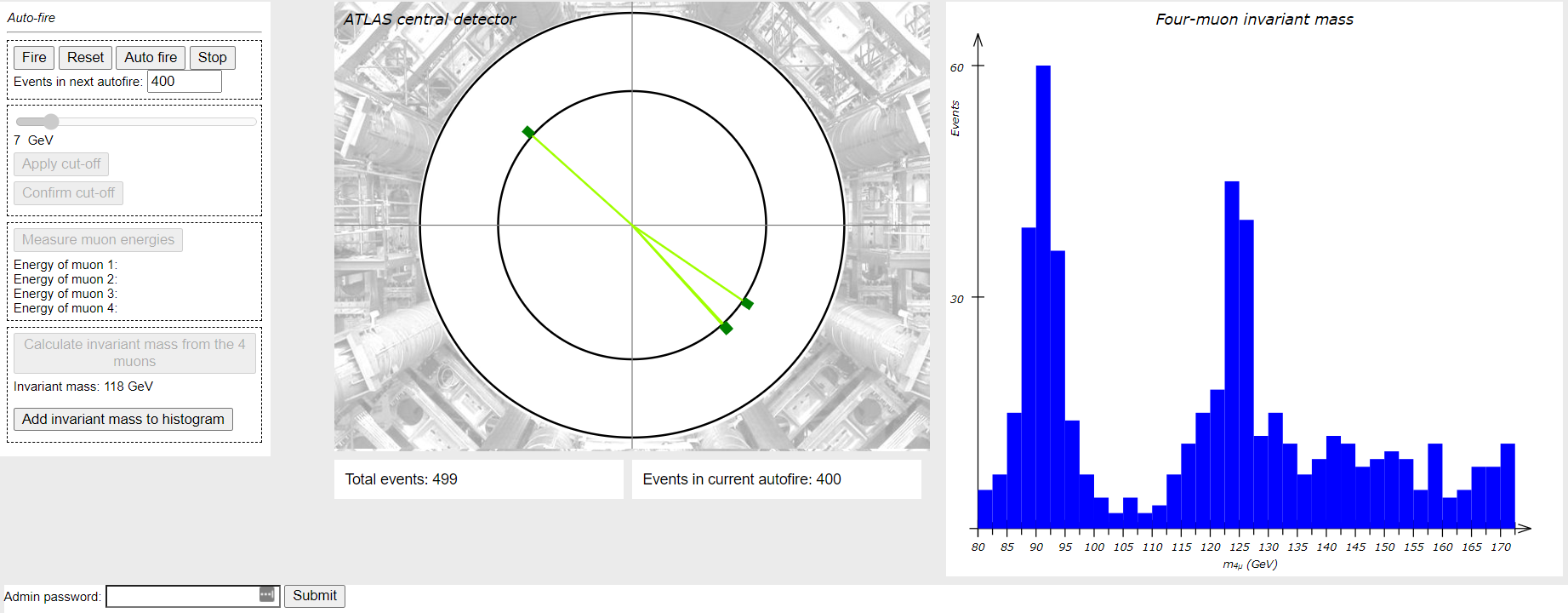}
        \caption{Screenshot of the interactive event display in the Higgs boson golden channel module.}
        \label{fig:Higgs_screenshot}
\end{figure}

The introduction page, user experience and admin page are similarly structured
as in the ALP with AFP module.
The interactive event display for this module is shown in Figure~\ref{fig:Higgs_screenshot}. 

\subsection{Sources of Publications}

The publication information is extracted using web scraping of the Fermilab web sites~\cite{CDF,D0} and 
with the CERN Document Server API~\cite{CDS}.
Publications of the following collaborations 
are included: 1989 – 2000: CERN – Large Electron-Positron Collider (LEP): ALEPH, DELPHI, L3, OPAL;  
1987 – 2011 Fermilab – Tevatron Collider: CDF, D0;
2010 – present: CERN – Large Hadron Collider (LHC):
ATLAS, CMS.

\subsection{Natural Language Processing}

An expert in the field of Higgs boson research recognizes immediately the category of an article.
However, this identification is not a 
trivial task for an automated computer categorization.
Therefore, a complex solution is chosen 
applying Artificial Intelligence (AI) and 
Natural Language Processing (NLP).
Details and further references are given in Ref.~\cite{Zacik:2774895}.

\subsection{User interface}
\begin{figure}[!hbtp]
    \centering
    \includegraphics[width=\textwidth]{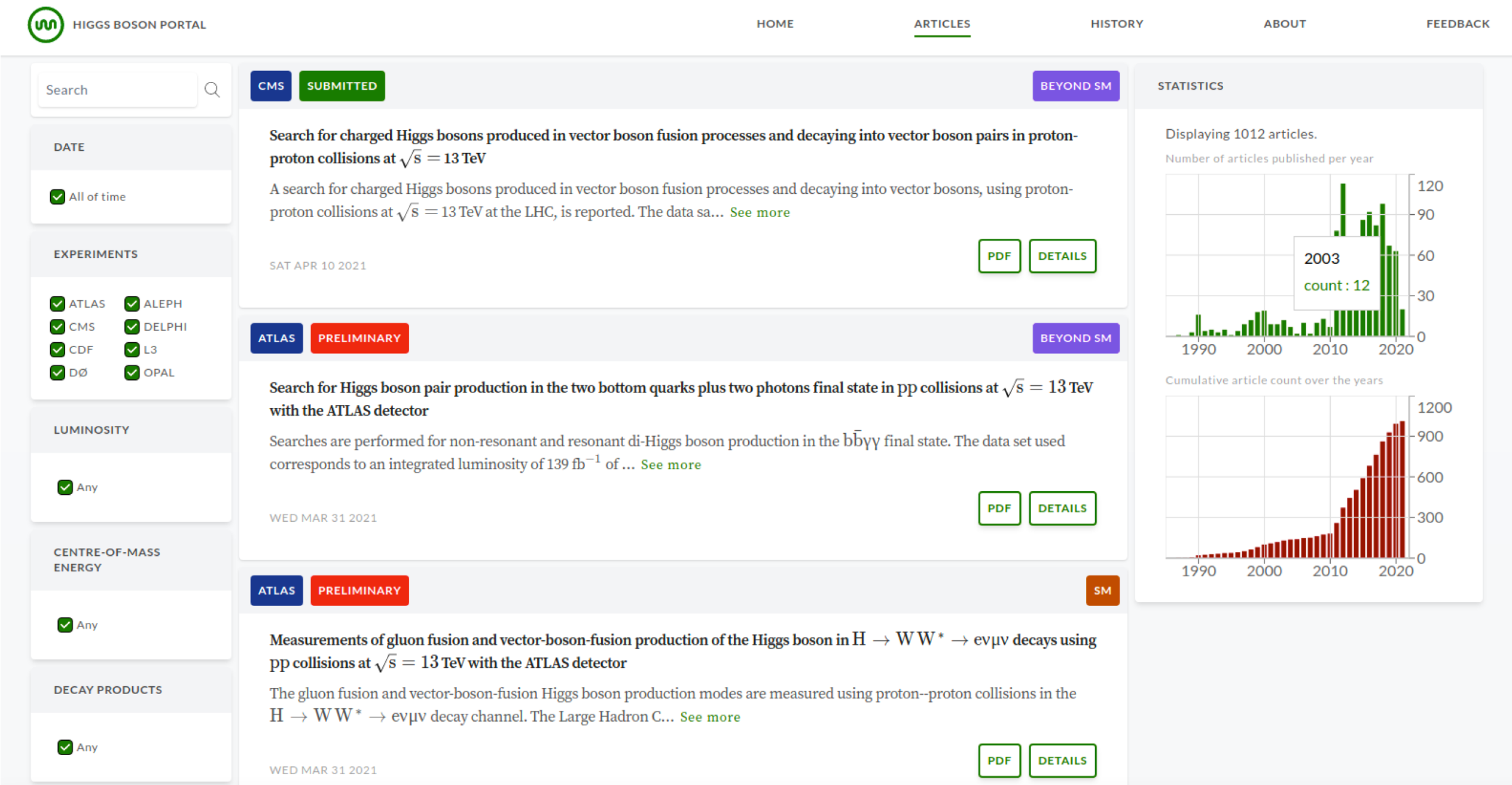}
    \vspace*{-7mm}
    \caption{User interface Higgs Boson Portal (HBP).}
    \label{fig:interface}
\end{figure}

The user interface is shown in Fig.~\ref{fig:interface}.

\subsection{Developments of Limits and Measurements}

\begin{figure}[!hbtp]
    \centering
    \includegraphics[width=0.32\textwidth]{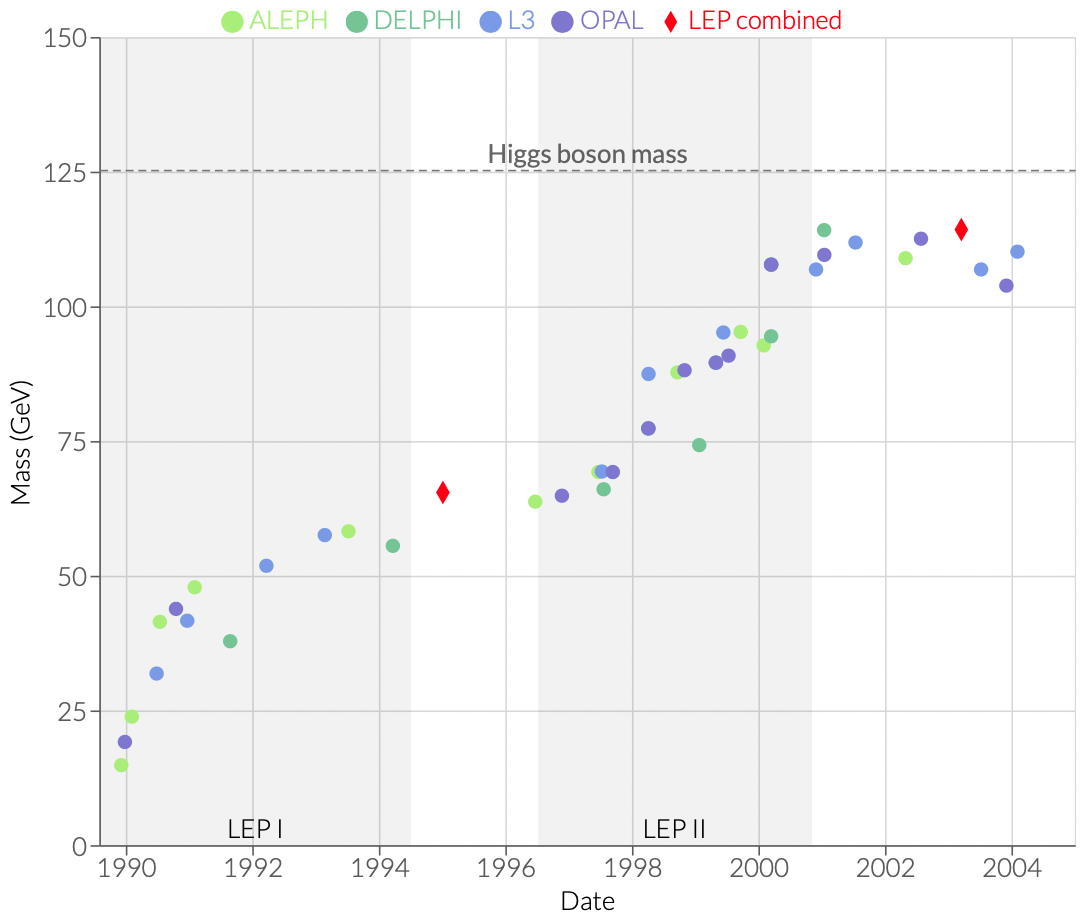}
    \includegraphics[width=0.32\textwidth]{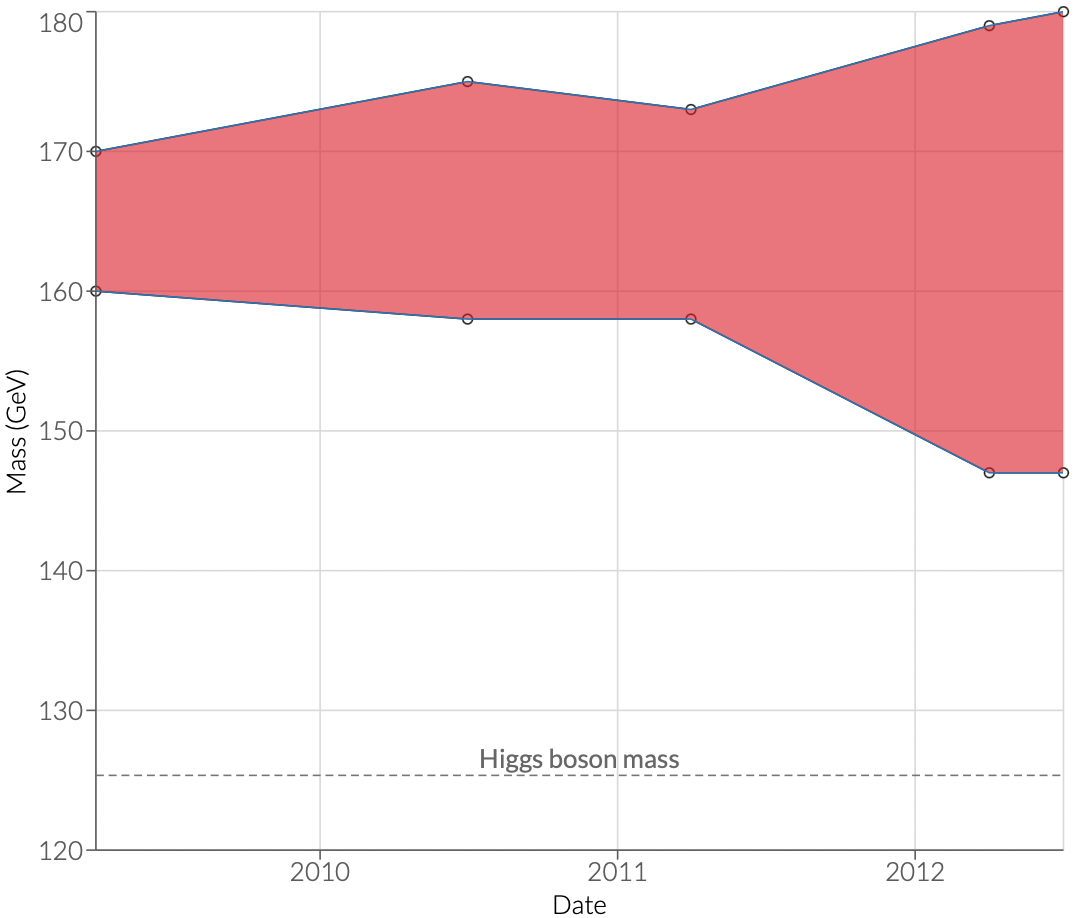}
 \includegraphics[width=0.32\textwidth]{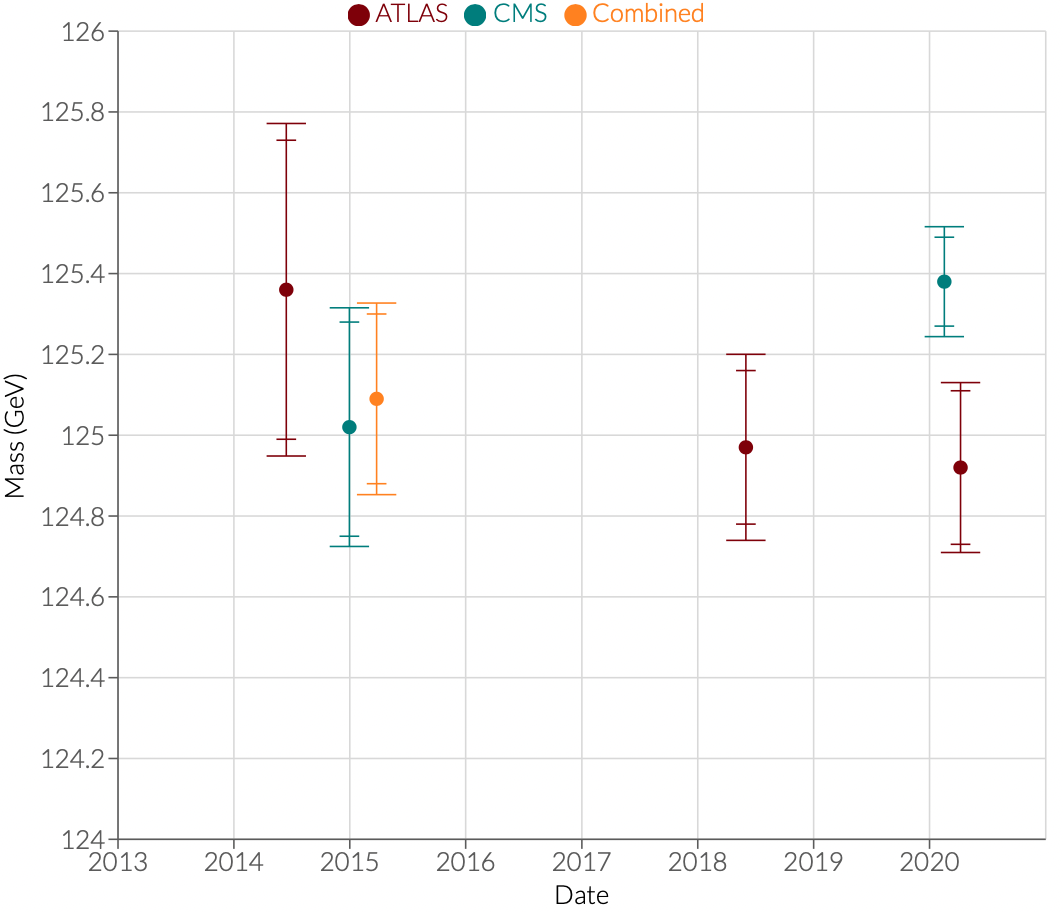}   
    \vspace*{-2mm}
    \caption{Developments of limits and 
measurements in the Higgs boson research.}
    \label{fig:history}
\end{figure}

The "History" link in the user interface gives 
examples of the developments of limits and 
measurements in the Higgs boson research (Fig.~\ref{fig:history}). Clicking on a data point leads to the corresponding publication, 
and the development of the upper Higgs boson 
mass limit is taken from Ref.~\cite{Sopczak_2012}.

\subsection{Administration and Maintenance}

The user interface allows administration login. 
The administrator can adjust the NLP categorization.
Categorized publications are stored in a PostgreSQL database.
Updates are performed daily with Python cron jobs.

\section{Conclusions} \label{sec:conclusions}

Two outreach modules have been developed and implemented for the Czech Particle Physics Project.
They are accessible at \href{http://cppp.web.cern.ch}{http://cern.ch/cppp},
and are ready to serve students, teachers, and the interested public.
Learning goals are:
giving school students 
a sense of the fun with science, technology, engineering and mathematics,
awakening interest in CERN, LHC and ATLAS physics,
understanding basic concepts in particle physics,
acquiring initial knowledge about elementary particles and how to detect them,
becoming aware that additional detectors help to separate signal and background events,
improving the understanding that increasing data statistics leads to new observations and discoveries.

In addition, Higgs Boson and Supersymmetry Portals have been established with more than 1000 categorized publications each and daily updates. They serve as an educational aid and data base for experts.

Feedback on the developed modules is very welcome and 
will be used in future updates of the modules. 
Further modules on related fields of research as well as 
modules on other research directions can be efficiently 
created and maintained using the structural setup in this project.

\vspace{-2mm}
\section*{Acknowledgements}
\vspace{-2mm}
The authors would like to 
thank the 
International Particle Physics Outreach Group
(IPPOG) for fruitful discussions.
The grant support from the "Fond celoškolských aktivit pro rok 2022+2023" Czech Technical University in Prague is gratefully acknowledged.
The project is also supported by the Ministry of
Education, Youth and Sports of the Czech
Republic under the project number
LTT 17018. A. Sopczak would like to acknowledge the DAAD support under project number 57705645.

\bibliographystyle{JHEP}
\apptocmd{\thebibliography}
 {\setlength{\parsep}{0pt}\setlength{\itemsep}{0pt plus 0.1pt}}
  {}{}
\bibliography{biblio}

\end{document}